\documentclass[prl,reprint,aps,amsmath,amssymb,twocolumn,superscriptaddress,longbibliography]{revtex4-1}
\usepackage{graphicx,bm,color,appendix}
\usepackage[colorlinks,bookmarks=true,citecolor=blue,linkcolor=blue,urlcolor=blue]{hyperref}
\usepackage{times}
\usepackage{amsmath,amssymb}
\usepackage{comment}
\usepackage{appendix}

\graphicspath{{./fig/}}

\usepackage{xcolor}
\definecolor{orange}{rgb}{1,0.5,0}

\newcommand{\tj}[6]{ \begin{pmatrix}
   #1 & #2 & #3 \\
   #4 & #5 & #6 
  \end{pmatrix}}

\usepackage[normalem]{ulem}
\usepackage{comment}

\begin{document}

\title{Operator Product Expansion Coefficients of the 3D Ising Criticality via Quantum Fuzzy Sphere }

\author{Liangdong Hu}
\affiliation{Westlake Institute of Advanced Study,	 Westlake University, Hangzhou 310024, China }

\author{Yin-Chen He}
\email{yhe@perimeterinstitute.ca}
\affiliation{Perimeter Institute for Theoretical Physics, Waterloo, Ontario N2L 2Y5, Canada}

\author{W. Zhu}
\email{zhuwei@westlake.edu.cn}
\affiliation{Westlake Institute of Advanced Study, Westlake University, Hangzhou 310024, China }

\begin{abstract}
Conformal field theory (CFT) is the key to various critical phenomena.
So far, most of studies focus on the critical exponents of various universalities, corresponding to conformal dimensions of CFT primary fields. However, other important yet intricate data such as the operator product expansion (OPE) coefficients governing the fusion of two primary fields, is largely unexplored before, specifically in dimensions higher than 2D (or equivalently $1+1$D). Here, motivated by the recently-proposed fuzzy sphere regularization, we investigate the operator content of 3D Ising criticality starting from a microscopic description.   
We first outline the procedure of extracting OPE coefficients on the fuzzy sphere, and then compute 13 OPE coefficients of low-lying CFT primary fields. 
The obtained results are in agreement with the numerical conformal bootstrap data of 3D Ising CFT within a high accuracy.
In addition, we also manage to obtain 4 OPE coefficients including $f_{T_{\mu\nu} T_{\rho\eta} \epsilon}$ that were not available before, which demonstrates the superior capabilities of our scheme.
By expanding the horizon of the fuzzy sphere regularization from the state perspective to the operator perspective, we expect a lot of new physics ready for exploration.
\end{abstract}

\date{\today}

\maketitle

Critical phenomena attract broad interests in various fields of physics, ranging from condensed matter to high energy physics. An intriguing reason is, enlarged symmetry often emerges at the low energy and large length scale in the vicinity of a critical point \cite{Sachdev_book,Cardy_book}. 
One well-developed scenario is that at a continuous phase transition where scale invariance is enhanced to the
conformal invariance \cite{polyakov1970conformal,polchinski1988scale,Dymarsky2015scale}, the low-energy physics is well captured by the conformal field theory (CFT) \cite{yellowbook,Belavin1984}, a field theory that is invariant under the conformal symmetry. 
The nature of a CFT, hence its corresponding critical point, is largely determined by a set of conformal data, consisting of a list of conformal dimensions and operator product expansion (OPE) coefficients \cite{OPE_Wilson_1969,OPE_Kadanoff_1969,polyakov1970conformal}. 
The complete conformal data is able to reproduce many universal properties of the phase transition and determine the stability of fluctuations close to a fixed point \cite{Cardy_book}.
For example, the scaling dimensions are critical exponents and OPE coefficients give rise to a universal (charge or thermal) conductivity~\cite{Katz2014} which are experimentally measurable at a critical point. 
It remains an outstanding challenge for various fields of physics to accurately determine the conformal data of interacting CFTs beyond 2D (or equivalently $1+1$D).

Generally speaking, there are two strategies to study the conformal data of an interacting critical system. 
The first strategy, adopted by the conformal bootstrap program \cite{RMP_CB}, utilizes general axioms of CFTs to constrain the conformal data.
Although the bootstrap approach can produce a great deal of conformal data accurately for certain critical theories such as the 3D Ising, it very often loses its power for many interesting universalities since its starting point is too general to make contact with a specific universality.
In contrast, an alternative strategy, adopted in condensed matter research, is to study the microscopic model that directly realizes the universality of interest \cite{Vicari2002}.
The weakness of this down-to-earth strategy is that, it can access only a very limited number of conformal data.
For example, in recent progress on computing OPE coefficients of 3D classical transitions, either directly from three-point correlators \cite{Hasenbusch2018,Hasenbusch2020}, or indirectly from off-critical two-point correlators \cite{OPE_MC_2015}, due to various complications very limited number of (only two or three) OPE coefficients were accessible in these computations~\footnote{One complication is, these simulations were done on the geometry $T^3$ or $T^2\times R$, for which the famous three-point conformal correlator is valid only when the distances $x_{ij}$ of operators are well separated from the UV (i.e. lattice spacing) and IR scale (system size), namely  $1\ll x_{ij} \ll L$.}. 

In the 1980s, Cardy has outlined a scheme to compute the conformal data in microscopic models by making use of the state-operator correspondence, i.e. radial quantization, on the cylinder geometry $S^{d-1}\times \mathbb R$ \cite{Cardy1984,Cardy1985}.
The OPE coefficients can be computed directly via the one-point expectation value $\langle \phi_\alpha | \phi_\beta |\phi_\gamma\rangle$.
This scheme not only greatly reduces computational complexities, but also in principle enables the access of all the conformal data that are infinitely many.
For the 2D (i.e. 1+1D) critical theory, this scheme has been applied extensively~\cite{CARDY1986,Blote1986Conformal,affleck1988universal,Zou2018,Zou2020,Zou2022} since one just needs to study a lattice quantum Hamiltonian defined on a circle ($S^1$, i.e. periodic boundary chain).
Moving to critical theories in higher dimensions, a fundamental obstacle arises because spatial spherical geometry $S^{m\ge 2}$ has a curvature such that any regular lattice cannot fit in. 
This fundamental obstacle has recently been removed by the two of us using an innovative idea, the \emph{fuzzy (non-commutative) sphere regularization}~\cite{ZHHHH2022}. 
The correspondence between scaling dimensions and energy spectra of the 3D Ising CFT has been demonstrated convincingly. In this Letter we are extending the horizon of the fuzzy sphere regularization to the precise determination of OPE coefficients. 
In particular, we have accurately computed 13 different OPE coefficients of the 3D Ising CFT,  including four OPE coefficients that were  unavailable in the existing literature. 
Our successful access of CFT operator contents in the 3D Ising criticality paves the way for studying higher dimensional CFTs through the fuzzy sphere regularization.

\textit{3D Ising CFT on the fuzzy sphere.--}We consider spinful electrons moving on the sphere $S^2$ in the presence of $4\pi\cdot s$ monopole placed in the sphere origin. 
Due to the monopole the single-electron states form quantized Landau levels, with $2s+1$ degenerate orbitals in the lowest Landau level (LLL) \cite{Sphere_LL_Haldane}.
In the case that the LLL is partially filled and the gap between the LLL and higher LLs is much larger than other interaction scales in the system, we can effectively project the system into the LLL.
After the LLL projection coordinates of electrons are not commuting any more, $ [\tilde x_\mu, \tilde x_\nu] =  i \frac{R}{s}   \epsilon_{\mu\nu\rho}   \tilde x_\rho$.
Thus, we end up with a system defined on a fuzzy (non-commutative) two-sphere \cite{madore1992fuzzy}.

In practice, the Hamiltonian on the fuzzy sphere can be written in the second quantized form using the basis of Landau orbitals.
Interestingly, these Landau orbitals form a spin-$s$ irreducible representation of $SO(3)$, and the sphere radius $R\sim \sqrt{s}$.
A 2+1D Ising transition can be realized by the following Hamiltonian~\cite{ZHHHH2022}
\begin{widetext}
\begin{align} \label{eq:HamIsing}
H & =  \sum_{m_{1,2,3,4}=-s}^s  V_{m_1,m_2,m_3,m_4} [\left(\mathbf{c}_{m_1}^\dag \mathbf{c}_{m_4} \right) 
\left( \mathbf{c}_{m_2}^\dag  \mathbf{c}_{m_3} \right) - \left(\mathbf{c}_{m_1}^\dag \sigma^z \mathbf{c}_{m_4} \right) \left( \mathbf{c}_{m_2}^\dag \sigma^z \mathbf{c}_{m_3} \right)]    -h \sum_{m=-s}^s \mathbf{c}_m^\dag \sigma^x \mathbf{c}_m,
\end{align}
\end{widetext}
where $\mathbf{c}^\dag_m = (c^\dag_{m\uparrow}, c^\dag_{m\downarrow})$ is the fermion creation operator on the $m_\textrm{th}$ Landau orbital and $\sigma^{x,y,z}$ is the Pauli matrix.
The element $V_{m_1,m_2,m_3,m_4}$ is connected to the Haldane pseudopotential $V_l$ \cite{Sphere_LL_Haldane} by
$V_{m_1, m_2, m_3, m_4} 
= \sum_l V_l \, (4s-2l+1)\tj{s}{s}{2s-l}{m_1}{m_2}{-m_1-m_2} 
\tj{s}{s}{2s-l}{m_4}{m_3}{-m_3-m_4}$,
where $\tj{j_1}{j_2}{j_3}{m_1}{m_2}{m_3}$ is the Wigner $3j$-symbol. 
In this paper we will only consider ultra-local  density-density interactions in real space, corresponding to non-zero Haldane pseudopotentials $V_0,V_1$. 

At the half-filling (i.e. electron number $N=2s+1$), the transverse field $h$ triggers a phase transition from a quantum Hall ferromagnet \cite{Girvin2000} with spontaneous $Z_2$ symmetry broken to a quantum paramagnet, which falls into the 2+1D Ising universality class. Hereafter, we consider the critical point at $V_1/V_0=4.75,h/V_0=3.16$ that is determined by the  order parameter scaling \cite{ZHHHH2022}. At the critical point, the state-operator correspondence, a one-to-one correspondence between Hamiltonian eigenenergies and the scaling dimensions of the CFT operators, has been convincingly demonstrated \cite{ZHHHH2022}.

\textit{Operators on the fuzzy sphere.--}One can appreciate the beauty of the fuzzy sphere regularization by contrasting it with the familiar lattice regularization/model: The former is defined in the continuum with the full space symmetry (i.e. $SO(3)$ rotation of sphere), while the latter is defined on a discrete lattice with only discrete symmetries such as the lattice translation and rotation.
More importantly, the continuum nature of the fuzzy sphere model does not lead to any UV divergence, we still have a finite dimensional Hilbert space to study, thanks to the representation in the Landau orbital basis \cite{ZHHHH2022,Ippoliti2018Half}.
We can easily translate between operators (and states) on the continuous sphere and those on the discrete Landau orbitals, e.g.
\begin{align}
\psi_a(\bm \Omega)^\dag & = \sum_{m=-s}^s c_{m,a}^\dag \, Y^{(s)}_{s,m} (\bm \Omega).  
\end{align}
Here we are using the spin-weighted spherical Harmonics (also called monopole Harmonics) $Y^{(s)}_{s,m} (\bm  \Omega)$  because of the monopole in the origin of the sphere. 
The more interesting operators are spin operators, which are the particle-hole pairs of electrons.
For example, the simplest spin operator is,
\begin{align}
n^a(\bm  \Omega) & = \psi^\dag(\bm  \Omega) \sigma^a \psi(\bm  \Omega) = N\sum_{l=0}^{2s} \sum_{m=-l}^l n^a_{l,m} Y_{l,m}(\bm  \Omega), \\ 
n^a_{l,m} &= \frac{1}{2s+1}\int d\bm  \Omega \,  \bar{Y}_{l,m}(\bm  \Omega) n^a(\bm  \Omega).
\end{align}
Here we are using spherical Harmonics $Y_{l,m}(\bm  \Omega)$ instead of spin-weighted spherical Harmonics  because the spin is not seeing the monopole placed in the origin. 
We note that the electronic charges are auxiliary degrees of freedom that are gapped, while spin degrees of freedom are gapless.

Any gapless spin operator $\hat{\mathcal O}(\bm  \Omega)$ in our microscopic model can be expressed as a linear combination of CFT scaling operators ($\hat \phi_\alpha (\bm  \Omega)$) including primaries and descendants,
\begin{equation} \label{eq:operator_expansion}
\hat{\mathcal O}(\tau=0,\bm  \Omega) = \sum_\alpha c_\alpha \hat \phi_\alpha (\tau=0, \bm  \Omega).
\end{equation}
$c_\alpha$ is a non-universal and operator dependent numerical factor, which is generically non-zero as long as the spin operator has same symmetry (i.e. Ising $\mathbb Z_2$ and parity symmetry) as the CFT scaling operator. 
Now we can use Eq.~\eqref{eq:operator_expansion} and cylinder correlators of CFT to compute OPE coefficients.
For any primary state $|\phi_\alpha\rangle$, $\langle \phi_\alpha| \hat \phi_\beta | 0\rangle$ is non-vanishing only if $\phi_\beta$ equals to $\phi_\alpha$ or its descendant, giving rise to \cite{SM}
\begin{equation} \label{eq:lattice2pt}
\langle \phi_\alpha| \mathcal  O(\tau=0, \bm \Omega) |0\rangle = \sum_{n=0}^\infty \frac{c_n h_n(\bm \Omega)}{R^{\Delta_\alpha+n}},
\end{equation}
where $h_n(\bm \Omega)$ is a universal function fixed by the conformal symmetry, and $h_n(\bm \Omega)=1$ if $\phi_\alpha$ and $\phi_\beta$ are both Lorentz scalar primaries \cite{SM}.
Similarly, we have
\begin{equation}\label{eq:OPEs}
\langle \phi_\alpha| \mathcal  O(\tau=0, \bm \Omega) |\phi_\gamma\rangle = \sum_{\beta} f_{\alpha\beta\gamma}\frac{c_\beta  \tilde{h}_{\alpha\beta\gamma}(\bm \Omega)}{R^{\Delta_\beta}}.
\end{equation}
Here $f_{\alpha\beta\gamma}$ is the universal OPE coefficients \cite{OPE_Wilson_1969,yellowbook}, as it determines the fusion channel of two primary operators $\lim_{\bm y\rightarrow \bm x} \phi_\alpha (\bm x) \times \phi_\beta(\bm y) \sim \sum f_{\alpha\beta\gamma} \phi_\gamma(\bm x)$. $\tilde{h}_{\alpha\beta\gamma}(\bm \Omega)$ is again a universal function  fixed by the conformal symmetry, and is identity if $\phi_{\alpha,\beta,\gamma}$ are Lorentz scalar primaries \cite{SM}. 
Here the summation is over all the primaries and descendants in the operator expansion Eq.~\eqref{eq:operator_expansion}.
Next we will demonstrate the method of extracting  OPE coefficients $f_{\alpha\beta\gamma}$ using Eq.\eqref{eq:lattice2pt}-\eqref{eq:OPEs}.

\begin{figure}[t]
\includegraphics[width=0.87\linewidth]{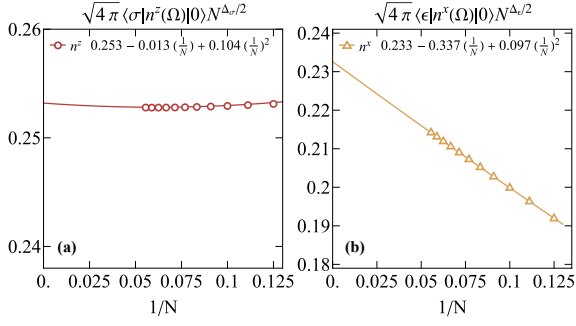}
\caption{\textbf{Local operator content}. 
 Extrapolation of (a) $\langle \sigma|\hat n^z(\mathbf \Omega) |0 \rangle \times N^{\Delta_\sigma/2}$, (b) $\langle \epsilon|\hat n^x(\mathbf \Omega) |0 \rangle \times N^{\Delta_\epsilon/2}$. The finite-size correction of $N^{-1}, N^{-2}$ respectively comes from the descendant fields as shown in Eq. \eqref{eq:lattice2pt}.
}
\label{fig:component}
\end{figure}

\textit{Extracting OPE coefficients.---}
To implement the idea of accessing OPE coefficients through spin operators, 
we need to firstly check if local spin operators produce the correct scaling in Eq.~\eqref{eq:lattice2pt}.
Below we will show the case for the $\mathbb Z_2$ odd operator $\hat n^z(\bm \Omega)$ and two $\mathbb Z_2$ even operators $\hat n^x(\bm \Omega)$ and $\hat O_\epsilon(\bm \Omega)\equiv \hat H(\bm \Omega) + 2 h \hat n^x(\bm \Omega) $, where $\hat H(\bm \Omega)$ is the Hamiltonian density, i.e. $\int d\bm \Omega \hat H(\bm \Omega)$ is our Hamiltonian in Eq.~\eqref{eq:HamIsing}.
Fig.~\ref{fig:component} shows the numerical data of candidate operators, in good agreement with the CFT expectation, indicating the lowest primary fields will dominate the observables of spin operators when the system size is large enough \footnote{
We recall that the electron number $N=2s+1$ plays the role of space volume, i.e. $N\sim R^2$. So in the numerical analysis we will just use $\sqrt{N}$ as the radius of sphere.}.  In particular, $\hat n^z(\bm \Omega)$ is dominated by $\hat \sigma$ (with conformal dimension $\approx 0.518$) with very small high order correction from its descendants, and $\hat n^x(\bm \Omega), \hat O_\epsilon(\bm \Omega)$ are dominant by $\hat \epsilon$ (conformal dimension $\approx 1.412$)~\footnote{$O_\epsilon(\bm \Omega)$ gives an identical result as $n^x(\bm \Omega)$ (up to a factor) as they differ by the Hamiltonian density.}.

\begin{figure}[t]
\includegraphics[width=0.98\linewidth]{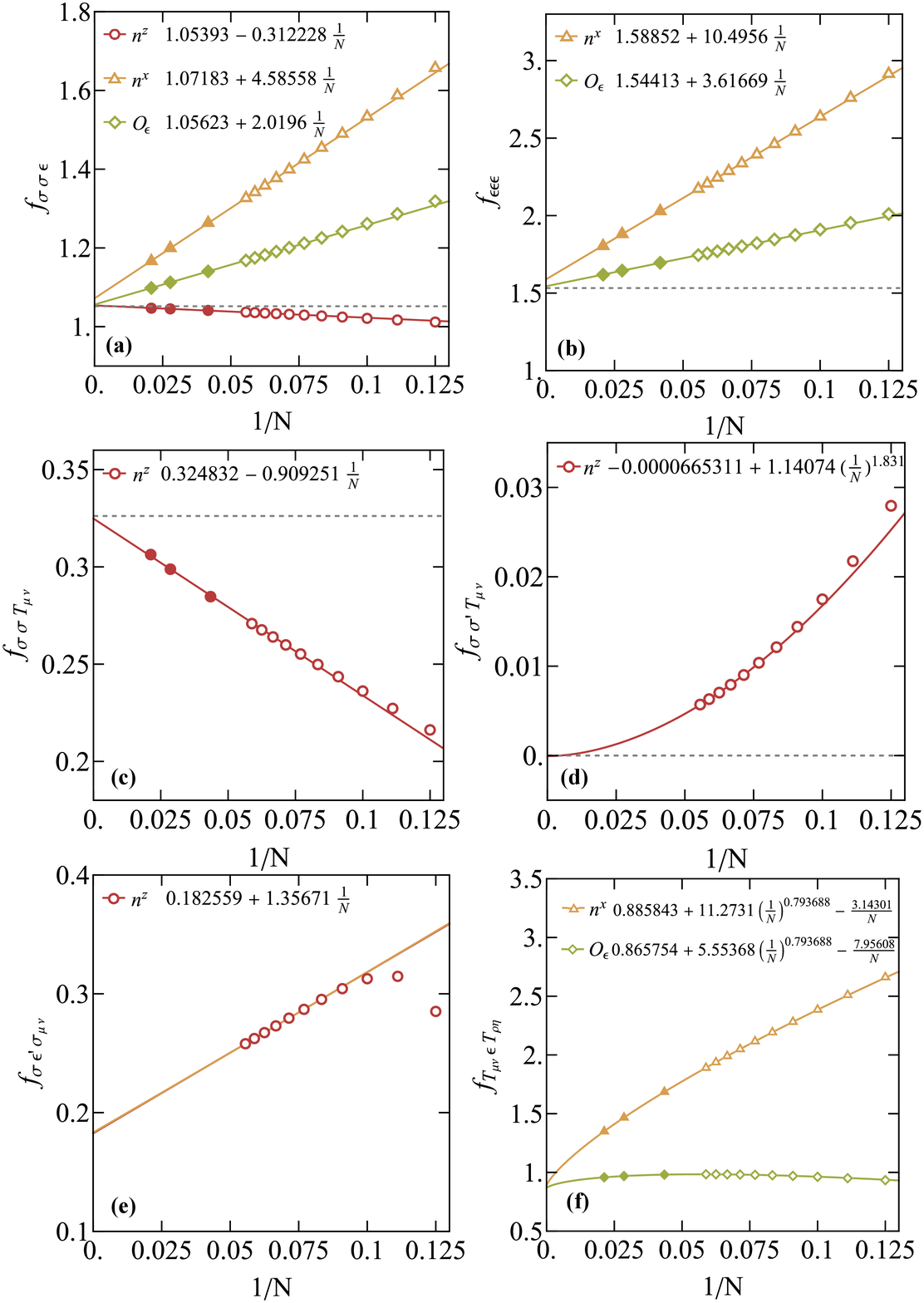}
\caption{\textbf{OPE coefficients of primary operators}.
(a-b) Two representative OPE coefficients involving three scalar primaries
 $f_{\sigma\sigma\epsilon}$ and $f_{\epsilon\epsilon\epsilon}$, obtained from the finite-size extrapolation via spin operator $n^z(\mathbf \Omega)$ (red), $n^x(\mathbf \Omega)$ (yellow) and $O_\epsilon(\mathbf \Omega)$ (green) (see main text). 
(c-d) Two representative OPE coefficients involving energy-momentum tensor $f_{\sigma\sigma T_{\mu\nu}}, f_{\sigma'\sigma T_{\mu\nu}}$ by using $n^z(\mathbf \Omega)$ operator.  
The dashed line in (a-d) is the value from conformal bootstrap. 
(e-f) Two representative OPE coefficients $f_{\sigma\epsilon'\sigma_{\mu\nu}} , f_{T_{\mu\nu}\epsilon T_{\mu\nu}} $ that are not known in conformal bootstrap calculation. In all figures only the data points on largest six sizes are used in the fitting.  
In (a-c) and (f), larger sizes up to $N=48$ are available by the DMRG method which are labeled by solid symbols. 
}
\label{fig:ope}
\end{figure}

Next we turn to extract the OPE coefficients.
Let us take the OPE coefficient  $f_{\sigma\sigma\epsilon}$ as an example.
As we explained above, to approximate $\hat \sigma$ a natural candidate is $n^z(\bm \Omega)$, and its operator expansion should naturally contain all the $\mathbb Z_2$ odd primaries $\sigma, \sigma_{\mu\nu}, \sigma', \cdots$ as well as their descendants $\partial_\mu \sigma, \square \sigma, \partial_\mu \sigma_{\mu\nu}, \cdots$. 
So we shall have,
\begin{align}
\langle \sigma | n^z(\bm \Omega) |0\rangle  &= \frac{1}{R^{\Delta_\sigma}}\left(c_\sigma + \sum_{n=1}^{\infty} \frac{a_n}{R^{2 n}} \right), \\
 \langle \sigma | n^z(\bm \Omega) |\epsilon\rangle &= \frac{f_{\sigma \sigma\epsilon}}{R^{\Delta_\sigma}}\left( c_\sigma +  \sum_{n=1}^{\infty} \frac{\tilde a_n}{R^{2 n}} \right) + ... 
\end{align}
where $'...'$ stands for the contribution from other primaries and associated descendants.
We note that only scalar scaling operators can contribute to these two observables.
Therefore, we can compute $f_{\sigma\sigma\epsilon}$ using 
\begin{equation}\label{eq:ope_sigmasigmaepsilon}
\frac{\langle \sigma | n^z(\bm \Omega) |\epsilon\rangle}{ \langle \sigma |  n^z(\bm \Omega) |0\rangle} = f_{\sigma\sigma\epsilon} + \frac{a_1 -\tilde a_1}{c_\sigma R^2} + O(R^{-4}).
\end{equation}
In principle, no finite size extrapolation is needed once the system size $R$ is large enough.
In the small size simulation like our case, we will perform a simple linear extrapolation with respect to $R^{-2}$ (i.e. $N^{-1}$) that includes the contribution from the descendant field $\square \sigma$.
Similarly, other OPE coefficient $f_{\sigma \phi_1 \phi_2}$ can be computed by the $\mathbb Z_2$ odd spin operator $n^z(\bm \Omega)$ using $f_{\sigma \phi_1 \phi_2}\approx \langle \phi_1| n^z(\bm \Omega) |\phi_2\rangle/\langle \sigma| n^z(\bm \Omega) |0\rangle$ and 
OPE coefficients $f_{\epsilon \phi_1 \phi_2}$ can be computed by the $\mathbb Z_2$ even spin operators $\hat n^x(\bm \Omega), \hat O_\epsilon(\bm \Omega)$. 
One should be cautious that certain OPE coefficient may have a different subleading term (see \cite{SM} for details).

First, we discuss the OPE coefficients involving three scalar primary operators.
For example, the OPE coefficient $f_{\sigma\sigma\epsilon}$ can be extracted by Eq. \eqref{eq:ope_sigmasigmaepsilon},
where the lowest correction comes from the descendant field $\square \sigma$.
Fig. \ref{fig:ope}(a) shows the finite-size extrapolation indeed agrees with Eq. \eqref{eq:ope_sigmasigmaepsilon}, giving $f_{\sigma\sigma\epsilon} \approx 1.0539$ by extrapolation. Alternatively, $f_{\sigma\sigma\epsilon}$ can be obtained by choosing spin operators 
$\mathcal{\hat O}(\bm \Omega) =n^x(\bm \Omega)$ and $\mathcal{\hat O}(\bm \Omega) = H(\bm \Omega)+2hn^x(\bm \Omega)$ to approximate the $\hat\epsilon$ primary (see Supple. Mat. Sec. III \cite{SM}). Fig.\ref{fig:ope}(a) confirms that different local operators give consistent estimates of $f_{\sigma\sigma\epsilon}$. 
Within the similar procedure, we also obtain a few more OPE coefficients involving scalar primary operators as listed in Tab. \ref{tab:ope}.

Second, for the OPE coefficients involving spinning operators, the calculation is largely parallel except that we need to consider the angle ($\bm \Omega$) dependence~\cite{SM}.  
The angle dependence can be eliminated by integrating over the sphere against the spherical Harmonics, for example, one representative OPE coefficient $f_{\sigma\sigma T_{\mu\nu}}$ should follow  (see \cite{SM}) 
\begin{align}
 \sqrt{\frac{15}{8}}\frac{\langle \sigma|\int d\bm \Omega \, \bar Y_{2,0}(\bm \Omega) \hat n^z(\bm \Omega)|T_{\mu\nu}, m=0\rangle}{\langle \sigma|\int d\bm \Omega \, \bar Y_{0,0}(\bm \Omega) \hat n^z(\bm \Omega) |0\rangle} &\approx  f_{\sigma\sigma T_{\mu\nu}} + \frac{a}{N}. 
\end{align}
The corresponding finite-size extrapolations are shown in Fig. \ref{fig:ope}(c) and the estimates of OPE are in Tab. \ref{tab:ope}. 

Third, we can calculate a few more OPE coefficients which are unknown before.  Fig. \ref{fig:ope}(e-f) show the results of $f_{\sigma\epsilon'\sigma_{\mu\nu}}$ and $f_{T_{\mu\nu} T_{\rho\eta}\epsilon}$, the latter is defined as
\begin{align} \label{eq:TTepsilon} 
& \frac{\langle T_{\mu\nu}, m=0|\int d\bm \Omega\, \hat {\mathcal{O}(\bm \Omega)} |T_{\mu\nu}, m=0\rangle-\langle 0| \int d \bm\Omega\, \mathcal O(\bm\Omega)|0\rangle}{\langle \epsilon|\int d\bm \Omega\, \hat {\mathcal{O}(\bm \Omega)}|0\rangle} \nonumber \\  &\approx f_{T_{\mu\nu} T_{\rho\eta}\epsilon}  + \frac{a}{N^{0.7937}} + \frac{b}{N}, 
\end{align}
where $\hat{\mathcal O} = \hat n^x, \hat O_\epsilon$. 
The estimated values are shown in Tab. \ref{tab:ope}~\footnote{We note that usually the OPE involving two or three spinning primaries could have more than one independent tensor structures (i.e. OPE coefficients), but  $f_{T_{\mu\nu} T_{\rho\eta}\epsilon}$ has only one independent tensor structure due to the conservation of $T_{\mu\nu}$~\cite{EMT_boot}. 
So it is enough to evaluate Eq.~\eqref{eq:TTepsilon}, but one may have to perform a basis transformation to compare $f_{T_{\mu\nu} T_{\rho\eta}\epsilon}$ computed in other (linear dependent) tensor structure.}.

To quantify the data we get, we compare these results with the numerical bootstrap data in Tab. \ref{tab:ope}. 
Most of OPE coefficients match known results from the conformal bootstrap \cite{RMP_CB,Ising_CB} with
discrepancies smaller than $2\%$. Importantly, we also compute four OPE coefficients which have not been computed before. 
The ability to access OPEs which are not known before shows the superiority of fuzzy sphere method. 
In addition, we present a systematical error analysis in Supple. Mat. \cite{SM}, where the mean values are estimated by $\hat n^z$ or $\hat O_\epsilon$, which have smaller finite-size effect, and the error bars are estimated by different fitting processes.

\begin{table}
\caption{List of OPE coefficients of primary operators obtained on the fuzzy sphere.
The primary operators in consideration have scaling dimensions $\Delta_{\sigma}\approx 0.51815$, $\Delta_{\epsilon}\approx 1.4126$, $\Delta_{\epsilon'}\approx 3.8297$, $\Delta_{\sigma'}\approx 5.2906$, $\Delta_{T_{\mu\nu}}=3$, and $\Delta_{\sigma_{\mu\nu}} \approx 4.1803$.
The conformal bootstrap (CB) data is from Ref. \cite{Ising_CB}, where some of unavailable data is labeled by `NA'.
We note that our convention for $f_{T_{\mu\nu} \epsilon T_{\rho\eta}}$ is $f_{T_{\mu\nu} \epsilon  T_{\rho\eta}}=\frac{1}{ 
 \sqrt{4\pi}} \int d\bm \Omega \,  \langle T_{\mu\nu}, m=0| \hat\epsilon(\bm \Omega)| T_{\rho\eta}, m=0\rangle$~\cite{SM}. }
\label{tab:ope}
\begin{center}
\begin{tabular}{c|c|c|c|c}
 \hline 
 \hline
 Operators & Spin & $Z_2$ &  $f_{\alpha\beta\gamma}$ (Fuzzy Sphere)  & $f_{\alpha\beta\gamma}$ (CB)  \\
 \hline 
 $\sigma$ & $0$ & $-$ &  $f_{\sigma\sigma\epsilon}\approx1.0539(18) $ & $f_{\sigma\sigma\epsilon} \approx1.0519 $  \\
 $\epsilon$ & $0$ & $+$ &  $f_{\epsilon\epsilon\epsilon} \approx 1.5441(23)$ & $f_{\epsilon\epsilon\epsilon} \approx1.5324 $  \\ 
 $\epsilon'$ & $0$ & $+$ &  $f_{\sigma\sigma\epsilon'}\approx0.0529(16)$ & $f_{\sigma\sigma\epsilon'} \approx0.0530 $  \\ 
 $ $ & $ $ &  $ $  & $f_{\epsilon\epsilon\epsilon'}\approx1.566(68)$  & $f_{\epsilon\epsilon\epsilon'} \approx1.5360 $\\ 
  $\sigma'$ & $0$ & $-$ &  $f_{\sigma'\sigma\epsilon}\approx0.0515(42) $ & $f_{\sigma'\sigma\epsilon}\approx0.0572 $  \\
 &  &    & $f_{\sigma'\sigma\epsilon'} \approx1.294(51) $ & NA  \\
 &  &    & $f_{\sigma'\epsilon\sigma'} \approx2.98(13) $  & NA\\
 $T_{\mu\nu}$ & $2$ & $+$ & $f_{\sigma\sigma T} \approx0.3248(35) $ & $f_{\sigma\sigma T} \approx0.3261 $  \\ 
   &  &    & $f_{\sigma'\sigma T} \approx-0.00007(96) $ & $f_{\sigma'\sigma T} = 0 $  \\ 
   &  &    & $f_{\epsilon\epsilon T} \approx0.8951(35) $  & $f_{\epsilon\epsilon T} \approx0.8892 $ \\ 
   &  &    & $f_{T\epsilon T} \approx 0.8658(69) $ & NA   \\ 
 $\sigma_{\mu\nu}$ & $2$ & $-$ &  $f_{\sigma\epsilon \sigma_{\mu\nu}} \approx0.400(33) $ & $f_{\sigma\epsilon \sigma_{\mu\nu}} \approx0.3892$  \\ 
  &  &  & $f_{\sigma\epsilon' \sigma_{\mu\nu}} \approx0.18256(69) $ & NA  \\ 
 \hline
 \hline
\end{tabular}
\end{center}
\end{table}

\textit{Summary and discussion.---}
We have explained the procedure of extracting the conformal data of emergent 3D conformal field theory (CFT) using the recently-proposed fuzzy sphere regularization.   By exploiting the state-operator correspondence, we accurately compute operator product expansion (OPE) coefficients of CFT primary fields with low-lying conformal dimensions in a $(2+1)$-D quantum Ising model. 
As far as we know, most of OPE coefficients reported here have never been studied in a microscopic model before. 
Some of OPE coefficients are not even computed by the conformal bootstrap calculations. 
More importantly, our work exposes the operator perspective of 3D CFTs on the fuzzy sphere, which enables a lot more important physics to explore.
For example, we can directly compute 3D CFT four-point correlators--the core object in the classic and modern story of conformal bootstrap. 
Another interesting direction is to study the operator growth of 3D CFTs under a quantum quench, which will lead to new insights on the quantum chaos of CFTs.
We foresee that the fuzzy sphere scheme will be a powerful tool to study quantum criticalities and 3D CFTs.

{\it Acknowledgments.---} We acknowledge the useful discussion with Yijian Zou. We thank C. Han, E. Huffman, J. Hofmann for collaboration on a related project. 
LDH and WZ were supported by National Natural Science Foundation of China (No.~92165102, 11974288) and the foundation of Westlake University.  
 Research at Perimeter Institute is supported in part by the Government of Canada through the Department of Innovation, Science and Industry Canada and by the Province of Ontario through the Ministry of Colleges and Universities.

\bibliography{cft_ope}

\onecolumngrid 
\newpage

\begin{center}
\textbf{\large Supplementary Material}
\end{center}
\setcounter{subsection}{0}
\setcounter{equation}{0}
\setcounter{figure}{0}
\renewcommand{\theequation}{S\arabic{equation}}
\renewcommand{\thefigure}{S\arabic{figure}}

	\appendix
	In this supplementary material, we will show more details to support the discussion in the main text. In Sec.~I, we discuss CFT correlators on the cylinder $S^2\times \mathbb R$. In Sec.~II, we clarify the definition of spin operators used in the main text. In Sec.~III, we show how to get the OPE coefficients using spin operators and derive the finite-size scaling forms. In Sec.~IV, we present an error analysis of the OPE coefficients.

	\section{I. Operator correlations in CFT}
	In this section, we will discuss correlators of CFT primary operators on the cylinder $S^{d-1}\times \mathbb R$, in particular using the relation of state-operator correspondence.
	\subsection{1. Correlators of scalar primaries}
	
	\subsubsection{a. $\langle 0|\hat\phi|\phi\rangle $}
	First, we consider the scalar-scalar correlation function on $\mathbb R^d$,
	\begin{equation}
	\langle0|\hat\phi(r,\Omega)\hat\phi(r',\Omega')|0\rangle = \frac{1}{(r^2+r'^2-2rr'\cos(\Omega-\Omega'))^\Delta},
	        \end{equation}
	where $(r,\Omega)$ are the spherical coordinates and $\Delta$ is the scaling dimension of field $\hat \phi$. Using the relation of state-operate correspondence 
 	\begin{equation}\label{seq:stateoperator}
	|\phi\rangle = \lim_{r\rightarrow 0}\hat\phi(r,\Omega)|0\rangle, \qquad \langle\phi| = \lim_{r\rightarrow\infty}r^{2\Delta}\langle0|\hat\phi(r,\Omega),
	\end{equation}
 we have
	\begin{equation}
	\langle 0|\hat\phi(r,\Omega)|\phi\rangle=\lim_{r'\rightarrow 0} \langle0|\hat\phi(r,\Omega)\hat\phi(r',\Omega')|0\rangle = \frac{1}{r^{2\Delta}}.
	        \end{equation}
	Then, we apply the Weyl-transformation $\tau =R \ln r $ to map $(r,\Omega)$ in $\mathbb R^d$ to $(\tau,\Omega)$ in $S^{d-1}\times\mathbb R$, where $R$ is the radius of $S^{d-1}$. The operator transforms as 
	\begin{equation}
	\hat \phi(r,\Omega)\rightarrow \hat \phi(\tau,\Omega) = \Lambda(r,\Omega)^{\Delta/2} \hat \phi(r, \Omega),
	        \end{equation}
	where $\Lambda = R^{-2}e^\frac{2\tau}{R}$ is the scale factor of this transformation and we can see it by considering the metric
	\begin{equation}
	ds^2 = dr^2+r^2d\Omega^2=R^{-2}e^\frac{2\tau}{R}\left( d\tau^2+R^2d\Omega^2  \right) = \Lambda(r,\Omega)\left( d\tau^2+R^2d\Omega^2  \right).
	        \end{equation}
	Substituting into the correlator and letting $\tau=0$, we finally get
	\begin{equation} \label{seq:twopoint_correlator}
		\langle 0|\hat\phi(\tau=0, \Omega)|\phi\rangle = \langle 0|\Lambda(r,\Omega)^{\Delta/2} \hat\phi(r, \Omega)|\phi\rangle\big|_{r=1} = \Lambda(r,\Omega)^{\Delta/2} \frac{1}{r^{2\Delta}}\big|_{r=1} = R^{-\Delta}.
	\end{equation}

	\subsubsection{b. $\langle \phi_1|\hat \phi_2|\phi_3\rangle$}
	Now we consider the scalar-scalar-scalar correlator
	\begin{equation}
	\langle 0|\hat\phi_i(r_i,\Omega_i)\hat\phi_j(r_j,\Omega_j)\hat\phi_k(r_k,\Omega_k)|0\rangle = \frac{f_{ijk}}{d_{ij}^{\Delta_i+\Delta_j-\Delta_k}d_{jk}^{-\Delta_i+\Delta_j+\Delta_k}d_{ik}^{\Delta_i-\Delta_j+\Delta_k}},
	        \end{equation}
	where $d_{12}=\sqrt{r_1^2+r_2^2-2r_1r_2\cos(\Omega_1-\Omega_2)}$ is the distance between $(r_1,\Omega_1)$ and $(r_2,\Omega_2)$ . Using state-operator correspondence relations Eq.~\eqref{seq:stateoperator} we have 
	\begin{equation}
	\langle \phi_i|\hat \phi_j(r_j,\Omega_j) |\phi_k\rangle = \lim_{r_i\rightarrow\infty,r_k\rightarrow0}
	r_i^{2\Delta_i}\langle 0|\hat\phi_i(r_i,\Omega_i)\hat\phi_j(r_j,\Omega_j)\hat\phi_k(r_k,\Omega_k)|0\rangle
	= \frac{f_{ijk}}{r_j^{-\Delta_i+\Delta_j+\Delta_k}}.
	        \end{equation}
	Similarly, under the Weyl-transformation
	\begin{equation}\label{seq:threepoint_correlator}
	\langle \phi_i|\hat \phi_j(\tau=0,\Omega)|\phi_k\rangle = f_{ijk} R^{-\Delta_j}.
        \end{equation}

Thus, one can compute the OPE coefficients through Eq. \ref{seq:twopoint_correlator} and Eq. \ref{seq:threepoint_correlator}
\begin{align}
  f_{ijk} = \frac{ \langle \phi_i|\hat \phi_j(\tau=0,\Omega)|\phi_k\rangle }{ \langle 0|\hat\phi(\tau=0, \Omega)|\phi\rangle }.
\end{align}

\subsection{2. Correlators of spinning primaries}
\label{sm:spin2}

	In this subsection, we consider correlators involving spinning primaries \cite{Costa2011,Kravchuk2016}.
 We start with two-point correlators of spinning primaries,
	\begin{equation}
		\langle O_{\Delta,\ell}(\bm r, \bm z_1) O_{\Delta,\ell}(0,\bm z_2)\rangle = \frac{(\bm z_1 \cdot \bm z_2 - 2 (\bm n \cdot \bm z_1)(\bm n \cdot \bm z_2) )^\ell}{r^{2\Delta}},
	\end{equation}
 where $\bm n = \frac{\bm r}{r}$.
For our interest we just consider $\ell=2$ while other $\ell$ can be analyzed in a similar fashion.
	Now we recover the indices by applying the differential operator in the auxiliary $z$-space, 
	\begin{equation}
		D_{z,\nu} = (D/2-1)\partial_{z_\nu} - z_\mu \partial_{z_\mu} \partial_{z_\nu} - \frac{1}{2} z_\nu \partial_{z_\mu} \partial_{z_\mu}.
	\end{equation}
	We have
	\begin{equation}
		\langle O_{\mu\nu} (\bm r) O_{\alpha\beta} (0)\rangle  =  \frac{3}{8} \frac{(3 \eta_{\alpha \nu} \eta_{\beta \mu} + 3 \eta_{\alpha \mu} \eta_{\beta \nu} -2 \eta_{\alpha\beta} \eta_{\mu\nu})}{r^{2\Delta}} - \frac{9}{4} \frac{\eta_{\alpha\mu} n_\beta n_\nu + \eta_{\alpha\nu} n_\beta n_\mu + \eta_{\beta\mu} n_\alpha n_\nu + \eta_{\beta\nu} n_\alpha n_\mu - 4 n_\alpha n_\beta n_\mu n_\nu}{r^{2\Delta}}
	\end{equation}
	In the spherical Harmonics basis, we have $O_{\ell=2, m=0}= N_{\ell=2,m=0} (2O_{33} - O_{11}-O_{22}) $.
	So $\langle O_{\ell=2, m=0} | O_{\ell=2, m=0} \rangle=1$ gives
	$N_{\ell=2,m=0} = \sqrt{2/27}$.
	Now we consider the scalar-scalar-spinning correlator (with explicit indices)
	\begin{equation}
		\lim_{r_1\rightarrow\infty,r_3\rightarrow 0} r_1^{2\Delta_1} \langle \phi_1(\bm r_1) \phi_2(\bm r) O_{\mu\nu} (\bm r_3) \rangle = \frac{\lambda_{123}}{2} \frac{3 n_{\mu} n_{\nu} - \eta_{\mu\nu}}{ r^{\Delta_2+\Delta_3-\Delta_1}}.
	\end{equation}
	So
	\begin{equation}
		\lim_{r_1\rightarrow\infty,r_3\rightarrow 0} r_1^{2\Delta_1} \langle \phi_1(\bm r_1) \phi_2(\bm r) O_{\ell =2, m=0} (\bm r_3) \rangle = \frac{\lambda_{123}}{2} \sqrt{2/27} \frac{-3 n_{1}^2 - 3 n_{2}^2 + 6n_{3}^2 }{ r^{\Delta_2+\Delta_3-\Delta_1}} =  \frac{\lambda_{123}}{\sqrt{6}} \frac{ -n_{1}^2 -  n_{2}^2 + 2n_{3}^2 }{ r^{\Delta_2+\Delta_3-\Delta_1}} .
	\end{equation}
	Here $\vec n = (\sin(\theta)\cos(\varphi), \sin(\theta)\sin(\varphi), \cos(\theta))$.
 Applying Weyl transformation and integrate the correlator against $ \bar Y_{l=2,m=0}(\Omega)$ on the cylinder, we have 
	\begin{equation}
		\int d\Omega \bar Y_{l=2,m=0}(\Omega) \langle \phi_1|  \phi_2(\tau=0, \Omega) |O_{\ell =2, m=0} \rangle = 2\sqrt{\frac{2\pi}{15}}  \lambda_{123} R^{-\Delta_2}
	\end{equation}
	Finally, we have 
	\begin{equation}\label{eq:sm_l2_mult}
		\frac{\int d\Omega \bar Y_{l=2,m=0}(\Omega) \langle \phi_1|  \phi_2(\tau=0,\Omega) |O_{\ell =2, m=0} \rangle}{ \int d\Omega \bar Y_{l=0,m=0}(\Omega) \langle \phi_2 | \phi_2(\tau=0,\Omega) | 0\rangle } = \sqrt{\frac{2}{15}} \lambda_{123}  = \sqrt{\frac{8}{15}} f_{123}
	\end{equation}

\clearpage
\section{II. Operators definition}
In this section, we give the definition of operators $\hat n^a(\Omega)$ and $\hat O_\epsilon(\Omega)$.

\subsection{1. Density operator $\hat n^a(\Omega)$}
The density operator is defined as 
\begin{equation}
	\hat n^a(\Omega) = \sum_{m_1,m_2} \bar{Y}_{s,m_2}^{(s_0)}(\Omega)Y_{s,m_1}^{(s_0)}(\Omega)
	\hat c^\dagger_{m_1}\sigma^a\hat c_{m_2}
\end{equation}
where $\sigma^a$ is $2\times2$ Pauli matrix and the notation $c^\dagger_{m_1}\sigma^a\hat c_{m_2}$ means
 $\sum_{\mu\nu}c^\dagger_{m_1,\mu}\sigma^a_{\mu\nu}\hat c_{m_2,\nu}$.
The angular momentum decomposition 
\begin{equation}
	\hat n^a(\Omega) = \sum_{m_1,m_2} \bar{Y}_{s,m_2}^{(s_0)}(\Omega)Y_{s,m_1}^{(s_0)}(\Omega)
	\hat c^\dagger_{m_1}\sigma^a\hat c_{m_2} = N\sum_{l,m}\hat n^a_{l,m}Y_{l,m}(\Omega)
\end{equation}
with
\begin{equation}\label{def:nlm}
	\begin{split}
		\hat n^a_{l,m} &= \frac{1}{2s+1}\int d\Omega \, \bar Y_{l,m}(\Omega)\hat n^a(\Omega)\\
		&= \sqrt{\frac{2l+1}{4\pi}}\sum_{m_1}(-1)^{3s+m_1+l}\tj{s}{l}{s}{-s}{0}{s} \tj{s}{l}{s}{-m_1}{m}{m_1-m} \hat c_{m_1}^\dagger \sigma^a\hat c_{m_1-m}\\
		&=\frac{1}{\sqrt{4\pi(2l+1)}}\sum_{m_1}(-1)^{3s+l+m+m_1}
		\langle s,s;s,-s|s,s;l,0 \rangle
		\langle s,m_1-m;s,-m_1|s,s;l,-m \rangle \hat c_{m_1}^\dagger \sigma^a \hat c_{m_1-m}.
	\end{split}
\end{equation}
Throughout this article, the notation $\hat n^a$ with $a=x,y,z,0$ means $\sigma^x,\sigma^y,\sigma^z,I$.

\subsection{2. $\hat{O}_\epsilon(\Omega)$ }

Here we define a new local operator 
\begin{equation}
\hat{O}_{\epsilon}(\Omega_a) = \int d \Omega_b U(\Omega_{ab}) [ n^0(\Omega_a) n^0(\Omega_b)- n^z (\Omega_a) n^z(\Omega_b)] + h n^x(\Omega_a).
\end{equation}
It is different from the Hamiltonian density by reversing the sign of transverse field. This operator is $\mathbb Z_2$ even, which is expected to have nonzero overlap with $\hat{\epsilon}$.

The spherical modes of $O(\Omega_a)$ is
\begin{align}
(\hat{O}_\epsilon)_{l,m} &= \int d\Omega_a\, \hat{O}(\Omega_a) \bar{Y}_{l,m} (\Omega_a) 
\end{align}

\clearpage
\section{III. OPE coefficients and finite-size scaling}
In this section, we will present a detailed analysis of OPE coefficients from the microscopic spin operators. Generally speaking, since spin operators we used are not the exact CFT primary fields, so the three-point correlators involve contributions from other primaries or descendants. Fortunately, we will show that many OPE coefficients can be extracted from the proper finite-size extrapolation.

We will choose local operator $\hat n^z(\Omega)$ to approach the CFT operator $\hat \sigma$, and $\hat n^x(\Omega)$, $\hat O_\epsilon(\Omega)$ to approach $\hat \epsilon$. 
Although CFT operators and spin operators are always local operators defined on the sphere, for computing OPE coefficients it is more convenient to use operators defined in the angular momentum (orbital) space, which are the spherical modes of the local operators, e.g.,
\begin{equation}
\hat O_{l,m} = \int d\Omega \, \bar Y_{l,m}(\Omega) \hat O(\Omega).
\end{equation}
So in the detailed analysis presented below, the computations are done in the orbital space.

\subsection{1.  $f_{\sigma\sigma\epsilon}$ }

The operator decomposition $n^z(\bm \Omega)$ generically is, 
\begin{align}\label{eq:sm_sigma_decom}
	\hat n^z(\bm \Omega) &= c_{\sigma}\hat \sigma(\bm \Omega) + c_{\partial_{\mu}\sigma} \partial_{\mu}\hat\sigma(\bm \Omega) + c_{\square\sigma}\square\hat \sigma(\bm \Omega)  + c_{\partial_{\mu}\partial_{\nu}\sigma} \partial_{\mu}\partial_{\nu}\hat\sigma(\bm \Omega)+\cdots\nonumber \\
 & + c_{\sigma_{\mu\nu}} \hat \sigma_{\mu\nu}(\bm \Omega) +  c_{\partial_\mu \sigma_{\mu\nu}} \partial_\mu \hat \sigma_{\mu\nu}(\bm \Omega) +  c_{\partial_\rho \sigma_{\mu\nu}} \partial_\rho \hat \sigma_{\mu\nu}(\bm \Omega) + \cdots  \nonumber \\ 
  &+ c_{\sigma'}\hat \sigma'(\bm \Omega)+ c_{\partial_{\mu}\sigma'} \partial_{\mu}\hat\sigma'(\bm \Omega) + c_{\square\sigma'}\square\hat \sigma'(\bm \Omega)  + c_{\partial_{\mu}\partial_{\nu}\sigma'} \partial_{\mu}\partial_{\nu}\hat\sigma'(\bm \Omega)+\cdots \nonumber \\
  & + \cdots
\end{align}
where each line represents components of a primary and its descendants, the conformal dimension of these operators are $\Delta_{\hat \sigma}\approx 0.51814$, $\Delta_{\hat \sigma'}\approx 5.2906$, and $\Delta_{\sigma_{\mu\nu}}\approx 4.1803$ and more other operators included in $\cdots$. 
Thus, the OPE coefficient can be extracted by $\frac{\langle \sigma|\hat n^z_{0,0}|\epsilon\rangle}{\langle \sigma|\hat n^{z}_{0,0}|0\rangle}$ for which only scalar scaling operators will contribute,
\begin{equation}\label{eq:sm_sigma_scaling}
\begin{split}
		\frac{\langle \sigma|\hat n^z_{0,0}|\epsilon\rangle}{\langle \sigma|\hat n^{z}_{0,0}|0\rangle} 
		&\approx 
		\frac{c_{\sigma} f_{\sigma\sigma\epsilon} R^{-\Delta_\sigma} + c_{\square\sigma} f_{\sigma,\square\sigma,\epsilon} R^{-(\Delta_\sigma+2)} + c_{\square^2\sigma} f_{\sigma,\square^2\sigma,\epsilon} R^{-(\Delta_\sigma+4)} + c_{\sigma'} f_{\sigma,\sigma',\epsilon} R^{-\Delta_{\sigma'}}}{ c_{\sigma}  R^{-\Delta_\sigma} + c_{\square\sigma}  R^{-(\Delta_\sigma+2)} + c_{\square^2\sigma}  R^{-(\Delta_\sigma+4)} + c_{\sigma'}  R^{-\Delta_{\sigma'}}  }\\
		&\approx  f_{\sigma\sigma\epsilon} + \frac{c_1}{R^2} +
	\frac{c_2}{R^{4}}+ O(R^{-4.77}) \approx f_{\sigma\sigma\epsilon} + \frac{c_1^\prime}{N} +
		\frac{c_2^\prime}{N^{2}} + O(N^{-2.38}) .
\end{split}
\end{equation}
In the last line of Eq.\eqref{eq:sm_sigma_scaling}, we change the variable from the spherical radius $R$ to the number of Ising spins $N$ on the fuzzy sphere. It shows that the OPE coefficients can be achieved by extrapolation $N\rightarrow \infty$.  

The OPE $\sigma\sigma\epsilon$ can also be defined as $\langle \sigma|\epsilon|\sigma\rangle$ and computed using a $Z_2$ even operator such as $\hat n^x(\bm \Omega)$
\begin{equation}\label{eq:sm_epsilon_decom}
	\hat n^x (\bm \Omega) = c_{I}\hat I + [c_{\epsilon}\hat \epsilon (\bm \Omega) + \cdots] + [c_{T_{\mu\nu}}\hat T_{\mu\nu} (\bm \Omega) + \cdots] + [c_{\epsilon'}\hat \epsilon' (\bm \Omega) + \cdots] + \cdots
\end{equation}
where each bracket refers to a primary and its descendants (labeled by $\cdots$), with conformal dimensions of the first few lying ones to be $\Delta_I = 0$, $\Delta_{\epsilon}\approx 1.4126$, $\Delta_{T_{\mu\nu}}=3$, $\Delta_{\epsilon'}\approx 3.8296$. The identity component should be subtracted, we have
\begin{equation}\label{eq:sm_epsilon_scaling}
	\begin{split}
		\frac{\langle \sigma|\hat n^x_{0,0}|\sigma\rangle-\langle 0|\hat n^x_{0,0}|0\rangle}{\langle \epsilon|\hat n^{x}_{0,0}|0\rangle} 
		&\approx
		\frac{c_{\epsilon} f_{\sigma\epsilon\sigma} R^{-\Delta_\epsilon} + c_{\square\epsilon} f_{\sigma,\square\epsilon,\sigma} R^{-(\Delta_\epsilon+2)} + c_{\epsilon'} f_{\sigma,\epsilon',\sigma} R^{-\Delta_{\epsilon'}} + c_{\square^2\epsilon} f_{\sigma,\square^2\epsilon,\sigma} R^{-(\Delta_\epsilon+4)}}{ c_{\epsilon}  R^{-\Delta_\epsilon} + c_{\square\epsilon} R^{-(\Delta_\epsilon+2)} + c_{\epsilon'} R^{-\Delta_{\epsilon'}} + c_{\square^2\epsilon} R^{-(\Delta_\epsilon+4)} } \\
		&\approx  f_{\sigma\sigma\epsilon} + \frac{c_1}{R^2} + \frac{c_2}{R^{2.4173}} + 
		\frac{c_3}{R^4} + O(R^{-4.4173}) \approx f_{\sigma\sigma\epsilon} + \frac{c_1^\prime}{N} +
		\frac{c_2^\prime}{N^{1.2087}} + O(N^{-2}) .
	\end{split}
\end{equation}
Similarly, the OPE $\sigma\sigma\epsilon$ can also be computed by local spin operator $\hat O_\epsilon(\bm \Omega)$. The finite-size scaling form is the same with different numerical factors $c_\alpha$.

\subsection{2. Scalar OPE coefficients: $f_{\epsilon\epsilon\epsilon}$, $f_{\sigma\sigma\epsilon'}$, $f_{\sigma'\sigma\epsilon}$, $f_{\epsilon\epsilon \epsilon'}$, $f_{\sigma'\epsilon \sigma'}$ and $f_{\sigma'\sigma\epsilon'}$ }
Similar to Eq. (\ref{eq:sm_epsilon_decom}-\ref{eq:sm_epsilon_scaling}), the finite-size scaling of OPE $\epsilon\epsilon\epsilon$ reads
\begin{equation}
\begin{split}
\frac{\langle \epsilon|\hat n^x_{0,0}|\epsilon\rangle - \langle 0|\hat n^x_{0,0}|0\rangle}{\langle \epsilon|\hat n^{x}_{0,0}|0\rangle} 
&\approx
\frac{c_{\epsilon} f_{\epsilon\epsilon\epsilon} R^{-\Delta_\epsilon} + c_{\square\epsilon} f_{\epsilon,\square\epsilon,\epsilon} R^{-(\Delta_\epsilon+2)} + c_{\epsilon'} f_{\epsilon,\epsilon',\epsilon} R^{-\Delta_{\epsilon'}} + c_{\square^2\epsilon} f_{\epsilon,\square^2\epsilon,\epsilon} R^{-(\Delta_\epsilon+4)}}{ c_{\epsilon}  R^{-\Delta_\epsilon} +  c_{\square\epsilon}  R^{-(\Delta_\epsilon+2)} + c_{\epsilon'}  R^{-\Delta_{\epsilon'}} + c_{\square^2\epsilon} R^{-(\Delta_\epsilon+4)} }\\
&\approx  f_{\epsilon\epsilon\epsilon} + \frac{c_1}{R^2} + \frac{c_2}{R^{2.4173}} + \frac{c_3}{R^4} + O(R^{-4.4173}) \approx  f_{\epsilon\epsilon\epsilon} + \frac{c_1^\prime}{N} +
\frac{c_2^\prime}{N^{1.2087}} + O(N^{-2}).
\end{split}
\end{equation}

Similarly,
\begin{gather}
	\frac{\langle \sigma|\hat n^z_{0,0}|\epsilon'\rangle}{\langle \sigma|\hat n^{z}_{0,0}|0\rangle} \nonumber
	\approx  f_{\sigma\sigma\epsilon'} +\frac{c_1}{R^2} +
	\frac{c_2}{R^{4}}+ O(R^{-4.77}) \approx f_{\sigma\sigma\epsilon'} + \frac{c_1^\prime}{N} +\frac{c_2^\prime}{N^{2}}+ O(N^{-2.38}) \\
	\frac{\langle \sigma'|\hat n^z_{0,0}|\epsilon\rangle}{\langle \sigma|\hat n^{z}_{0,0}|0\rangle} 
	\approx  f_{\sigma'\sigma\epsilon} + 
	\frac{c_1}{R^2} +
	\frac{c_2}{R^{4}}+ O(R^{-4.77}) \approx f_{\sigma'\sigma\epsilon} + \frac{c_1^\prime}{N} +\frac{c_2^\prime}{N^{2}}+ O(N^{-2.38}) \\
	\frac{\langle \sigma'|\hat n^x_{0,0}|\sigma\rangle}{\langle \epsilon|\hat n^x_{0,0}|0\rangle} \nonumber
	\approx  f_{\sigma'\sigma\epsilon} + \frac{c_1}{R^2} + \frac{c_2}{R^{2.4173}} +
	\frac{c_3}{R^4} + O(R^{-4.4173}) \approx f_{\sigma'\sigma\epsilon} + \frac{c_1^\prime}{N} + \frac{c_2^\prime}{N^{1.2087}}+O(N^{-2}), \\
 \frac{\langle \epsilon|\hat n^x_{0,0}|\epsilon'\rangle}{\langle \epsilon|\hat n^x_{0,0}|0\rangle} \nonumber
	\approx  f_{\epsilon\epsilon\epsilon'} + \frac{c_1}{R^2} + \frac{c_2}{R^{2.4173}} +
	\frac{c_3}{R^4} + O(R^{-4.4173}) \approx f_{\epsilon\epsilon\epsilon'} + \frac{c_1^\prime}{N}+\frac{c_2^\prime}{N^{1.2087}} + O(N^{-2})\\
\end{gather}
In  the last two equations, the identity component shouldn't be subtracted since $\langle \sigma'|\hat I|\sigma\rangle = \langle \epsilon'|\hat I|\epsilon\rangle = 0$.
The numerical results are shown in Fig.\ref{fig:ope_sm1}(a-b), (e) and can be fitted as(up to linear term)
\begin{equation}
\begin{split}
	\frac{\langle \sigma|\hat n^z_{0,0}|\epsilon'\rangle}{\langle \sigma|\hat n^{z}_{0,0}|0\rangle} 
	&\approx 0.0529389-\frac{0.35087}{N}\\
	\frac{\langle \sigma'|\hat n^z_{0,0}|\epsilon\rangle}{\langle \sigma|\hat n^{z}_{0,0}|0\rangle} 
	&\approx 0.0514531-\frac{0.329505}{N}\\
	\frac{\langle \sigma'|\hat n^x_{0,0}|\sigma\rangle}{\langle \epsilon|\hat n^x_{0,0}|0\rangle}
	&\approx 0.052771-\frac{0.66503}{N} \\ 
 \frac{\langle \epsilon|\hat n^x_{0,0}|\epsilon'\rangle}{\langle \epsilon|\hat n^x_{0,0}|0\rangle}
	&\approx 1.56597-\frac{4.10764}{N} 
\end{split}
\end{equation}
where the extracted OPE coefficient are $f_{\sigma\sigma\epsilon'}\approx0.0529389$, $f_{\sigma'\sigma\epsilon}\approx0.0514531(\hat n^z),0.052771(\hat n^x)$ and $f_{\epsilon\epsilon \epsilon'} \approx 1.56597$. The conformal bootstrap results are $f^{CB}_{\sigma\sigma\epsilon'}\approx0.053012(55), f^{CB}_{\sigma'\sigma\epsilon}\approx0.057235(20)$, and $f_{\epsilon\epsilon \epsilon'}^{CB} \approx 1.5360(16)$.

Finally, we compute two OPE coefficients $f_{\sigma'\epsilon\sigma'}$ and $f_{\sigma'\sigma\epsilon'}$ that have not been computed by conformal bootstrap so far. We have,
\begin{equation}
\begin{split}
\frac{\langle \sigma'|\hat n^x_{0,0}|\sigma'\rangle-\langle 0|\hat n^x_{0,0}|0\rangle}{\langle \epsilon|\hat n^x_{0,0}|0\rangle} &
	\approx  f_{\sigma'\epsilon\sigma'} + \frac{c_1}{R^2} + \frac{c_2}{R^{2.4173}} +
	\frac{c_3}{R^4} + O(R^{-4.4173}) \approx f_{\sigma'\epsilon\sigma'} + \frac{c_1^\prime}{N} +\frac{c_2^\prime}{N^{1.2087}}+ O(N^{-2})\\
	\frac{\langle \sigma'|\hat n^z_{0,0}|\epsilon'\rangle}{\langle \sigma|\hat n^{z}_{0,0}|0\rangle} 
	&\approx  f_{\sigma'\sigma\epsilon'} +\frac{c_1}{R^2} +
	\frac{c_2}{R^{4}}+ O(R^{-4.77}) \approx f_{\sigma'\sigma\epsilon'} + \frac{c_1^\prime}{N}  +\frac{c_2^\prime}{N^{2}}+ O(N^{-2.38}).
\end{split}
\end{equation}
The numerical results are shown in Fig.\ref{fig:ope_sm1}(f-g). 
The fitting results of $f_{\sigma'\epsilon \sigma'}$ are 
	\begin{gather*}
		\frac{\langle \sigma'|\hat n^x_{0,0}|\sigma'\rangle-\langle 0|\hat n^x_{0,0}|0\rangle}{\langle \epsilon|\hat n^x_{0,0}|0\rangle}
		\approx 3.17844+\frac{30.9552}{N} \\
		\frac{\langle \sigma'|(\hat O_\epsilon)_{0,0}|\sigma'\rangle-\langle 0|(\hat O_\epsilon)_{0,0}|0\rangle}{\langle \epsilon|(\hat O_\epsilon)_{0,0}|0\rangle}
		\approx 2.97839+\frac{5.66358}{N},
	\end{gather*}
And the result of $f_{\sigma'\sigma \epsilon'}$ is 
$$ \frac{\langle \sigma'|\hat n^z_{0,0}|\epsilon'\rangle}{\langle \sigma|\hat n^z_{0,0}|0\rangle}
	\approx 1.29367-\frac{3.05227}{N}. $$

\subsection{3. OPE coefficients with one spinning operator:  $f_{\sigma\sigma T_{\mu\nu}}$, $f_{\sigma'\sigma T_{\mu\nu}}$, $f_{\epsilon\sigma\sigma_{\mu\nu}}$,  $f_{\epsilon'\sigma\sigma_{\mu\nu}}$, and $f_{\epsilon\epsilon T_{\mu\nu}}$}
The OPE coefficients involving spinning operator is slightly more complicated, since one has to carefully deal with the $\bm \Omega$ dependence.
As derived in Sec. I.2 we can compute $\int d\Omega \, \bar Y_{l=2,m=0}(\Omega) \langle \sigma|\hat n^z (\Omega)|T_{\mu\nu}, m=0\rangle=\langle \sigma|\hat n^z_{2,0}|T_{\mu\nu},m=0\rangle$ for which all the scalar and $\ell=2$ scaling operators will contribute, so
\begin{equation}\label{eq:sm_sigma_spin2_scaling}
\begin{split}
\sqrt{\frac{15}{8}}\frac{\langle \sigma|\hat n^z_{2,0}|T_{\mu\nu}\rangle}{\langle \sigma|\hat n^{z}_{0,0}|0\rangle} 
&\approx 
\frac{c_{\sigma} f_{\sigma\sigma T_{\mu\nu}} R^{-\Delta_\sigma} + c_{\square\sigma} f_{\sigma,\square\sigma,T_{\mu\nu}} R^{-(\Delta_\sigma+2)} + c_{\sigma_{\mu\nu}} f_{\sigma,\sigma_{\mu\nu},T_{\mu\nu}} R^{-\Delta_{\sigma'}} + c_{\square^2\sigma} f_{\sigma,\square^2\sigma,T_{\mu\nu}} R^{-(\Delta_\sigma+4)}}{ c_{\sigma}  R^{-\Delta_\sigma} + c_{\square\sigma}  R^{-(\Delta_\sigma+2)} + c_{\square^2\sigma}  R^{-(\Delta_\sigma+4)} + c_{\sigma'}  R^{-\Delta_{\sigma'}}  } \\
&\approx  f_{\sigma\sigma T_{\mu\nu}} + \frac{c_1}{R^2} + \frac{c_2}{R^{3.662}} + O(R^{-4}) \approx f_{\sigma\sigma T_{\mu\nu}} + \frac{c_1^\prime}{N} + \frac{c_2^\prime}{N^{1.831}} + O(N^{-{2}}).
\end{split}
\end{equation}

\begin{figure}[t]
	\includegraphics[width=0.98\linewidth]{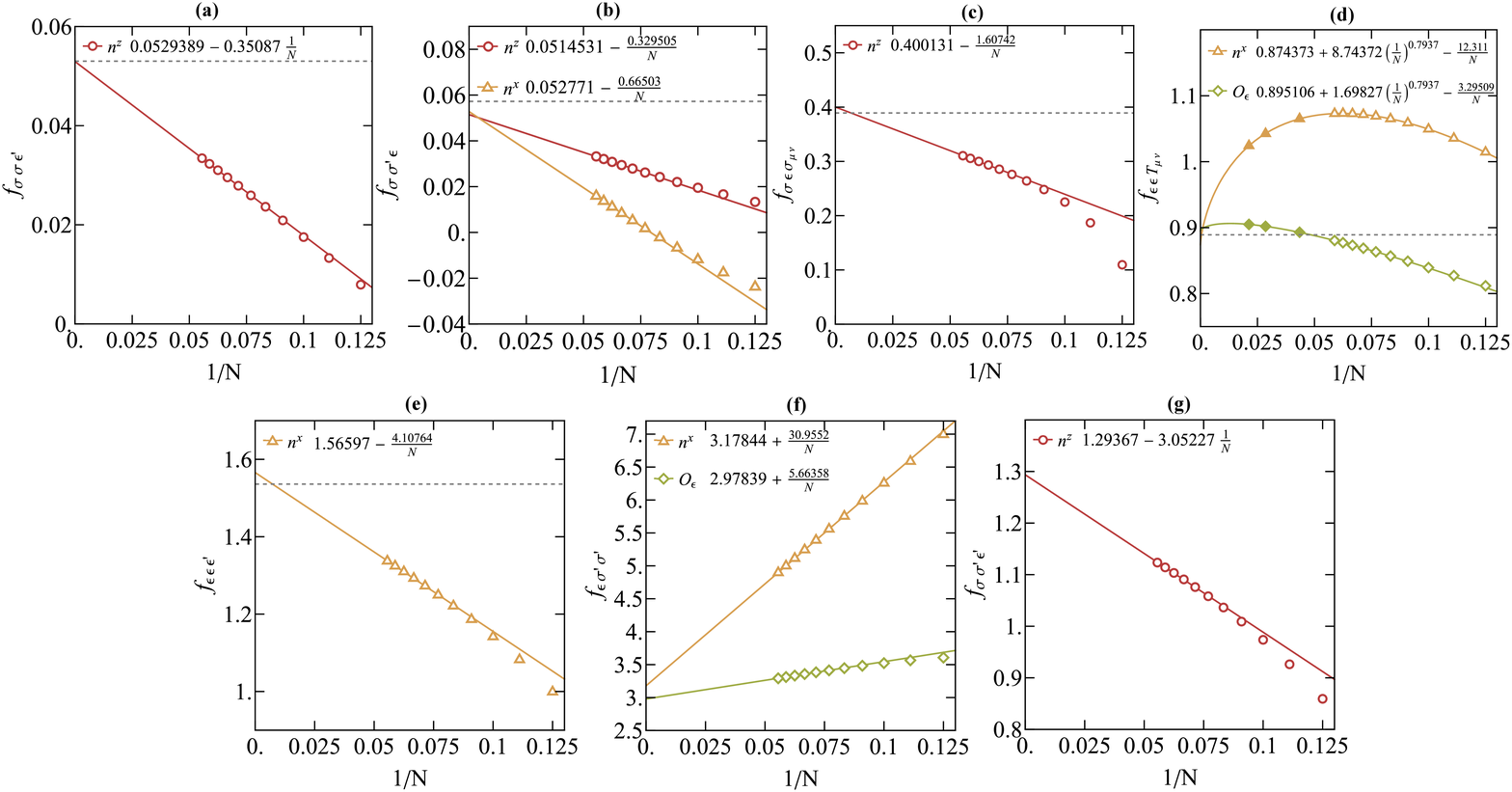}
	\caption{\textbf{More OPE coefficients of primary operators}. 	
        Finite-size extrapolation of OPE coefficients (see discussion in Sec. III 7-11).
		The dashed line in (a-e) is the value from conformal bootstrap ($f^{CB}_{\sigma\sigma\epsilon'}\approx0.053012(55), f^{CB}_{\sigma'\sigma\epsilon}\approx0.057235(20), f^{CB}_{\epsilon\sigma\sigma_{\mu\nu}}\approx0.38915941(81)$ and $f^{CB}_{\epsilon\epsilon T_{\mu\mu}}\approx0.8891471(40)$, $f_{\epsilon\epsilon \epsilon'}^{CB} \approx 1.5360(16)$). Only the data points from $N=13$ to $18$ ($13$ to $48$ if DMRG data is available) are used in the fitting. The last two OPE(f-g) have not been studied in conformal bootstrap.
	}
	\label{fig:ope_sm1}
\end{figure}

$f_{\sigma'\sigma T_{\mu\nu}}$, $f_{\epsilon\sigma\sigma_{\mu\nu}}$, $f_{\epsilon'\sigma\sigma_{\mu\nu}}$, and $f_{\epsilon\epsilon T_{\mu\nu}}$ can be computed in a similar fashion.
In specific we have,
\begin{equation}
\begin{split}
\sqrt{\frac{15}{8}}\frac{\langle \sigma'|\hat n^z_{2,0}|T_{\mu\nu}\rangle}{\langle \sigma|\hat n^{z}_{0,0}|0\rangle} 
&\approx  f_{\sigma'\sigma T_{\mu\nu}} + \frac{c_1}{R^{3.662}} +
\frac{c_2}{R^{4.772}} + O(R^{-5.662}) \approx f_{\sigma'\sigma T_{\mu\nu}} + \frac{c_1^\prime}{N^{1.831} } + \frac{c_2^\prime}{N^{2.386} } + O(N^{-{2.831}}).
\end{split}
\end{equation}
where the leading correction comes from $f_{\sigma'\sigma_{\alpha\beta}T_{\mu\nu}}$.
Similarly,
\begin{equation}
\sqrt{\frac{15}{8}} \frac{\langle \epsilon|\hat n^z_{2,0}|\sigma_{\mu\nu}\rangle}{\langle \sigma|\hat n^{z}_{0,0}|0\rangle} 
\approx  f_{\epsilon \sigma \sigma_{\mu\nu}} + \frac{c_1}{R^2} + \frac{c_2}{R^{3.662}} + \frac{c_3}{R^4} + O(R^{-4.772}) \approx f_{\epsilon \sigma \sigma_{\mu\nu}} + \frac{c_1^\prime}{N} + \frac{c_2^\prime}{N^{1.831}}  + O(N^{-2}).
\end{equation}
The numerical results are shown in Fig.\ref{fig:ope_sm1}(c) and its fitting result is
$$ \sqrt{\frac{15}{8}} \frac{\langle \epsilon|\hat n^z_{2,0}|\sigma_{\mu\nu}\rangle}{\langle \sigma|\hat n^{z}_{0,0}|0\rangle} \approx
	0.400131-\frac{1.60742}{N}. $$
The extracted OPE coefficient is $f_{\epsilon\sigma\sigma_{\mu\nu}}\approx0.400131$ which is closed to conformal bootstrap result $f^{CB}_{\epsilon\sigma\sigma_{\mu\nu}}\approx0.38915941(81)$.

For $f_{\epsilon'\sigma\sigma_{\mu\nu}}$ we have
\begin{align}
\sqrt{\frac{15}{8}}  \frac{\langle \epsilon'|\hat n^z_{2,0}|\sigma_{\mu\nu}\rangle }{\langle \sigma|\hat n^{z}_{0,0}|0\rangle} 
&\approx  f_{\epsilon'\sigma\sigma_{\mu\nu}} + \frac{c_1}{R^2} + \frac{c_2}{R^{3.662}}+\frac{c_3}{R^4} + O(R^{-4.772}) \approx f_{\epsilon'\sigma\sigma_{\mu\nu}} + \frac{c_1^\prime}{N} +
\frac{c_2^\prime}{N^{1.831}} + O(N^{-2}).
\end{align}

The OPE coefficient $f_{\epsilon\epsilon T_{\mu\nu}}$ is slightly different as we are using a $\mathbb Z_2$ even operator for the computation, 
\begin{equation}
	\begin{split}
	\sqrt{\frac{15}{8}} 	\frac{\langle \epsilon|\hat n^x_{2,0}|T_{\mu\nu}\rangle}{\langle \epsilon|\hat n^{x}_{0,0}|0\rangle} 
		&\approx f_{\epsilon\epsilon T_{\mu\nu}} + \frac{c_1}{R^{1.5874}} + \frac{c_2}{R^2} + 
		\frac{c_3}{R^{3.5874}} + O(R^{-4}) \approx f_{\epsilon\epsilon T_{\mu\nu}} + \frac{c_1^\prime}{N^{0.7937}} +
		\frac{c_2^\prime}{N}+\frac{c_3^\prime}{N^{1.7937}} + O(N^{-2}) .
	\end{split}
\end{equation}
Here the leading correction ($R^{-1.5874}$) comes from the $T_{\mu\nu}$ component contained in $n^x(\bm \Omega)$. 
The numerical results are shown in Fig.\ref{fig:ope_sm1}(d), we fit the data up to linear term
\begin{gather*}
\sqrt\frac{15}{8}\frac{\langle \epsilon|\hat n^x_{2,0}|T_{\mu\nu}\rangle}{\langle \epsilon|\hat n^{x}_{0,0}|0\rangle}
\approx   0.874373+\frac{8.74372}{N^{0.7937}}-\frac{12.311}{N} \nonumber\\
\sqrt\frac{15}{8}\frac{\langle \epsilon|(\hat O_\epsilon)_{2,0}|T_{\mu\nu}\rangle}{\langle \epsilon|(\hat O_\epsilon)_{0,0}|0\rangle}
\approx   0.895106+\frac{1.69827}{N^{0.7937}}-\frac{3.29509}{N}.
\end{gather*}
The conformal bootstrap result is $f_{\epsilon\epsilon T_{\mu\nu}}^{CB}\approx0.8891471(40)$.

\subsection{4. $f_{T_{\mu\nu} \epsilon 
 T_{\rho\eta}}$}
The finite-size scaling of $f_{T_{\mu\nu} \epsilon T_{\rho\eta}}$ is similar to  $\epsilon\epsilon\epsilon$ 
\begin{equation}
\begin{split}
	\frac{\langle T_{\mu\nu}, m=0| \mathcal O| T_{\rho\eta}, m=0\rangle-\langle 0| \mathcal O|0\rangle}{\langle \epsilon| \mathcal O|0\rangle} \nonumber
	&\approx  f_{T_{\mu\nu} \epsilon  T_{\rho\eta}}  + \frac{c_1}{R^{1.5874}} + \frac{c_2}{R^2} + 
		\frac{c_3}{R^{2.4173}} + O(R^{-3})\\
  &\approx f_{T_{\mu\nu} \epsilon  T_{\rho\eta}} + \frac{c_1^\prime}{N^{0.7937}} +
		\frac{c_2^\prime}{N} +\frac{c_3^\prime}{N^{1.2087}}+ O(N^{-2}) .
\end{split}
\end{equation}
where $\mathcal O$ is chosen to be $n^x_{0,0}, (\hat O_\epsilon)_{0,0}$.
Therefore, the OPE we are computing is defined as $$f_{T_{\mu\nu} \epsilon  T_{\rho\eta}}=\frac{1}{\sqrt{4\pi}} \int d\bm \Omega \,  \langle T_{\mu\nu}, m=0| \hat\epsilon(\bm \Omega)| T_{\rho\eta}, m=0\rangle,$$
with  $$|T_{\mu\nu}, m=0\rangle= \lim_{r\rightarrow 0} \sqrt{\frac{2}{27}} (2T_{33}(\bm r) - T_{11}(\bm r)-T_{22}(\bm r)) |0\rangle.$$
 
\section{IV. Error analysis}
In this section, we analyze the error of obtained OPE coefficients. The stragety used to estimate errors are explained below:

First, we fit all data up to the lowest order according to the finite-size scaling functions as shown in Sec. III. 
This process gives the mean values of OPE coefficients $\bar{f}_{\alpha\beta\gamma}(\mathcal O)$ using local operators $\hat{\mathcal{O}}=\hat n^z$ or $\hat O_\epsilon$. In specific, for the OPE involving $\hat \sigma$ field like $f_{\alpha\beta\sigma}$, we use the operator $\hat n^z$ to estimate the mean value, while 
for the OPE involving $\hat \epsilon$ field  
we estimate $f_{\alpha\beta \epsilon} $ using $\hat{O}_\epsilon$ operator. (We have confirmed that $\hat n^x$ operator gives consistent results as $\hat{O}_\epsilon$ operator, but $n^x$ operator shows much larger finite-size dependence than that of $\hat O_\epsilon$. So we use $\hat O_\epsilon$ to estimate the OPE invloving $\epsilon$ field, and use $\hat n^x$ operator as a consistency check.  )

Second, we repeat all data fitting up to the second lowest order in the finite-size corrections  as shown in Sec. III.. 
The obtained estimates of OPE coefficients are denoted as 
$q_{\alpha\beta\gamma}(\hat{\mathcal{O}})$.

Third, the errors of OPE coefficients are estimated by $\Delta f_{\alpha\beta\gamma} \equiv |\bar{f}_{\alpha\beta\gamma}(\mathcal O) - q_{\alpha\beta\gamma}(\mathcal O)|$. 
Using above methods, the mean values and associated errors ($\bar{f}_{\alpha\beta\gamma} \pm \Delta f_{\alpha\beta\gamma}$) are listed in Tab. I in the main text. 

\end{document}